\documentclass[12pt]{article}
\pdfoutput=1
\usepackage{amsmath,amssymb,epsfig,ulem}
\usepackage[bookmarks=false]{hyperref}
\usepackage{color}
\usepackage{multicol}

\allowdisplaybreaks



\newcommand{\beq}{\begin{equation}}
\newcommand{\eeq}{\end{equation}}
\newcommand{\bea}{\begin{eqnarray}} 
\newcommand{\eea}{\end{eqnarray}}

\begin{document}


\begin{flushright}
MITP/13-043\\
PITT-PACC-1306
\end{flushright}

\vspace{15pt}
\begin{center}
  \Large\bf 
Charge Asymmetry in Top Pair plus Jet Production\\
-- A Snowmass White Paper --
\end{center}
\vspace{5pt}
\begin{center}
{\sc Stefan Berge$^{a}$ and Susanne Westhoff$^{b}$}\\
\vspace{10pt} {\sl
$^{a}$ PRISMA Cluster of Excellence, Institut f\"ur Physik (WA THEP), \\
Johannes Gutenberg-Universit\"at,
D-55099 Mainz, Germany} \\
\vspace{5pt} {\sl
$^b$ PITTsburgh Particle physics, Astrophysics \& Cosmology Center (PITT PACC), \\
Department of Physics and Astronomy, University of Pittsburgh, Pittsburgh, PA 15260, USA} \\
\end{center}
\vspace{5pt}
\begin{abstract}
\vspace{2pt} 
\noindent
We investigate the discovery potential of the top-quark charge asymmetry at the LHC in top-antitop production in association with a hard jet. In this process, the charge asymmetry can be accessed via two novel observables: the {\it incline asymmetry}, which probes the quark-antiquark channel, and the {\it energy asymmetry}, which gives access to the quark-gluon channel. At $8\,\text{TeV}$ collision energy, the significance for both observables is statistically limited. With $14\,\text{TeV}$ and an integrated luminosity of $50\,\text{fb}^{-1}$ or more, an asymmetry of up to $-12\%$ can be observed with a statistical significance of more than $5$ standard deviations. Prospects of measuring the charge asymmetry at the intended high-luminosity and high-energy LHC upgrades are discussed.
\end{abstract}

\section{Incline and energy asymmetries}
The persistence of the enhanced forward-backward asymmetry in top-antitop production observed at the Tevatron \cite{Aaltonen:2012it,Abazov:2011rq} calls for a cross check at the LHC experiments. For this purpose, it is necessary to define observables that are appropriate for the different experimental environment. In general, the measurement of a charge asymmetry in proton-proton collisions at high energies is complicated by the large symmetric background from gluon-gluon fusion, as well as by the symmetry of the initial state. In a first approach, the ATLAS and CMS collaborations have analyzed the charge asymmetry in terms of absolute rapidity differences, $A_C^{|y|}$ \cite{ATLAS:2012an,Chatrchyan:2012cxa}. No hint of any deviation from the standard-model prediction has been observed. However, due to the smallness of this observable, current analyses struggle with statistical and systematic uncertainties. A dedicated Snowmass study~\cite{Agashe:2013} suggests that even with high luminosity at the LHC with $14$ TeV collision energy (LHC14), the discovery of a charge asymmetry via $A_C^{|y|}$ remains challenging. It will depend critically on the extent to which systematic uncertainties can be reduced with increased luminosity.

So far, all attempts to measure the top-quark charge asymmetry have focused on inclusive top-antitop production. Yet, the exclusive mode with an additional hard jet in the final state provides us with new options to define observables. This process is abundant at the LHC, where a large number of top-antitop events are accompanied by one or more hard jets. In QCD, the charge asymmetry in $t\bar t + j$ production arises at leading order through the interference of initial- and final-state radiation. The corresponding rapidity asymmetries have been investigated up to next-to-leading order (NLO)~\cite{Dittmaier:2008uj,Melnikov:2010iu}. Those observables do not take account of the jet kinematics in the final state. Since the jet momentum is tied to the top and antitop momenta via momentum conservation, integrating over the jet kinematics reduces the sensitivity to the charge asymmetry in most regions of the phase space. Thus, in a recent work, we have elaborated on two novel observables that make use of the jet kinematics in order to maximize the accessibility of the charge asymmetry \cite{Berge:2013xsa}.

In the quark-antiquark channel, the charge asymmetry is preferably probed by the incline asymmetry $A^{\varphi,q}$. It is based on the inclination angle $\varphi$ between the planes spanned by initial- and final-state momenta in the partonic 
center-of-mass (CM) frame.\footnote{For the partonic channel $q\bar{q}\to t\bar{t}g$, $\varphi$ is defined by $\cos\varphi  =  \vec{n}_{qg}\cdot\vec{n}_{tg}$ with 
$\vec{n}_{qg} = (\vec{k}_q\times\vec{k}_g)/|\vec{k}_q\times\vec{k}_g|$ and $\vec{n}_{tg} = (\vec{k}_t\times\vec{k}_g)/|\vec{k}_t\times\vec{k}_g|$.
For the general definition of $\varphi$ and other quantities, as well as for a detailed discussion of the incline and energy asymmetries, we refer the reader to \cite{Berge:2013xsa}.} Since this inclination angle is per definition perpendicular to the jet direction, the impact of the jet kinematics on the charge asymmetry is minimized. At the LHC, the incline asymmetry is defined by
\begin{eqnarray}
A^{\varphi,q} & = & \frac{\sigma_A^{\varphi}(y_{t\bar t j} > 0) - \sigma_A^{\varphi}(y_{t\bar t j} < 0)}{\sigma_S}\,,\quad \sigma_A^{\varphi} = \int_0^{\pi}\text{d}\theta_j\left[\frac{\text{d}\sigma(\cos\varphi > 0)}{\text{d}\theta_j}-\frac{\text{d}\sigma(\cos\varphi < 0)}{\text{d}\theta_j}\right],
\end{eqnarray}
where $\sigma_S$ is the total cross section in $t\bar t + j$ production and $\theta_j$ is the jet scattering angle off the incoming quark in the partonic CM frame. The boost of the final state along the beam axis, $y_{t\bar tj}$, helps to determine the direction of the incoming quark, which is needed to reconstruct the inclination angle $\varphi$.

In the quark-gluon channel, the charge asymmetry can be accessed through the energy asymmetry $A^E$. It is defined in terms of the difference $\Delta E$ between the top and antitop energies in the partonic CM frame,
\begin{eqnarray}
A^E = \frac{\sigma(\Delta E > 0) - \sigma(\Delta E < 0)}{\sigma_S},\qquad \Delta E = E_t - E_{\bar t}\,.
\end{eqnarray}
Notice that $A^E$ does not depend on the direction of the incoming quark, which facilitates its observation at a symmetric hadron collider. The energy asymmetry and the incline asymmetry are to be considered as complementary observables. They probe the same origin of the charge asymmetry in terms of Feynman diagrams, but in different kinematical regions.

\section{Observability at LHC8 and LHC14}
In order to measure the charge asymmetry at the LHC, it is indispensable to enhance the signal strength by appropriate cuts. In a first step, we apply detector cuts on the  jet's transverse momentum $p_T^j \ge 25$ GeV and its rapidity in the laboratory frame $|y_j| \le 2.5$. 
To increase the absolute value of $A^{\varphi,q}$ and $A^E$, three further types of cuts prove useful: A lower cut on the boost of the final state, $|y_{t\bar t j}|$, enhances valence quark contributions with respect to the symmetric gluon-gluon initial state; an upper cut on the jet's rapidity in the partonic CM frame, $|\hat{y}_j|$, rejects the region of collinear jet emission; and lower limits on $|\cos\varphi|$ and $|\Delta E|$ further emphasize the kinematical features of the respective asymmetry. The strength and combination of these cuts needs to be optimized in order to maximize the signal-to-background ratio for a given amount of data.

At a collision energy of $8$ TeV (LHC8), the asymmetries can reach values up to $A^{\varphi,q} = -6\%$ and $A^E = -14\%$ in QCD at leading order for strong cuts (and correspondingly low production rates) \cite{Berge:2013xsa}. The enhanced maximum of $A^E$ compared to $A^{\varphi,q}$ is due to the larger abundance of quark-gluon rather than quark-antiquark initial states. However, with the data set of $22\,\text{fb}^{-1}$ collected by each experiment, the observability of $A^{\varphi,q}$ and $A^E$ is statistically limited. Assuming an experimental efficiency of $5\%$, an incline asymmetry of $A^{\varphi,q} = -2.4\,\%$ is accessible with a statistical significance of $3.6$ standard deviations for loose cuts. For the energy asymmetry, the maximum statistical significance amounts to $3.3\,\sigma$, corresponding to $A^E = -1.9\,\%$. The main limiting factor is the gluon-gluon background, whose suppression requires tight cuts and therefore larger event samples. A combination of the data collected by ATLAS and CMS is welcome to enhance the statistical sensitivity to the charge asymmetry at LHC8. 

\begin{figure}[!t]
\begin{center}
\includegraphics[height=7.7cm]{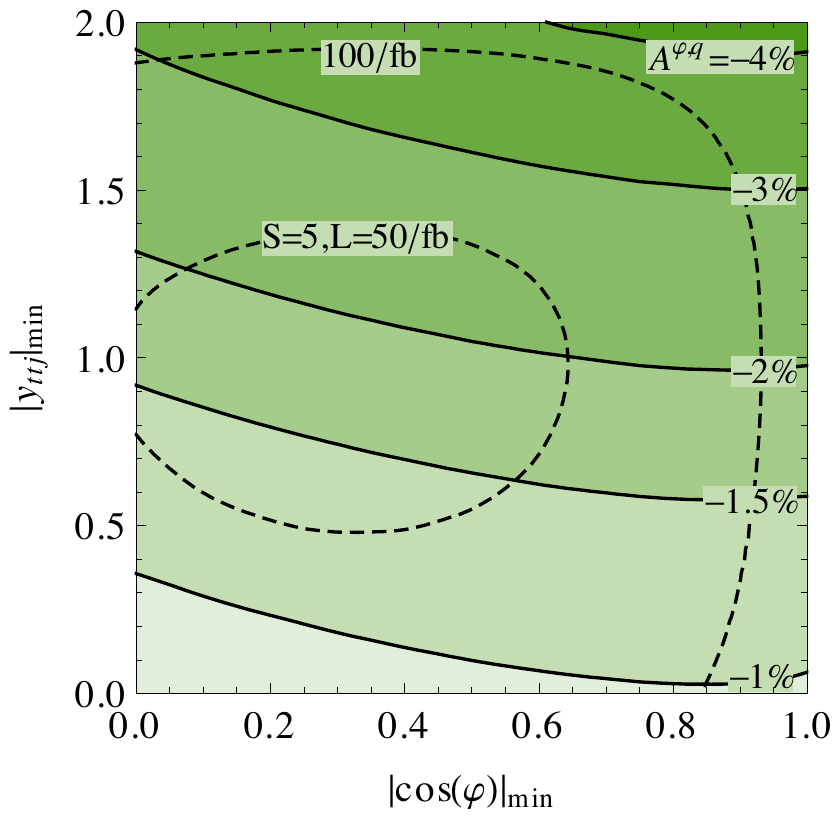}
\hspace*{0.4cm}
\includegraphics[height=7.7cm]{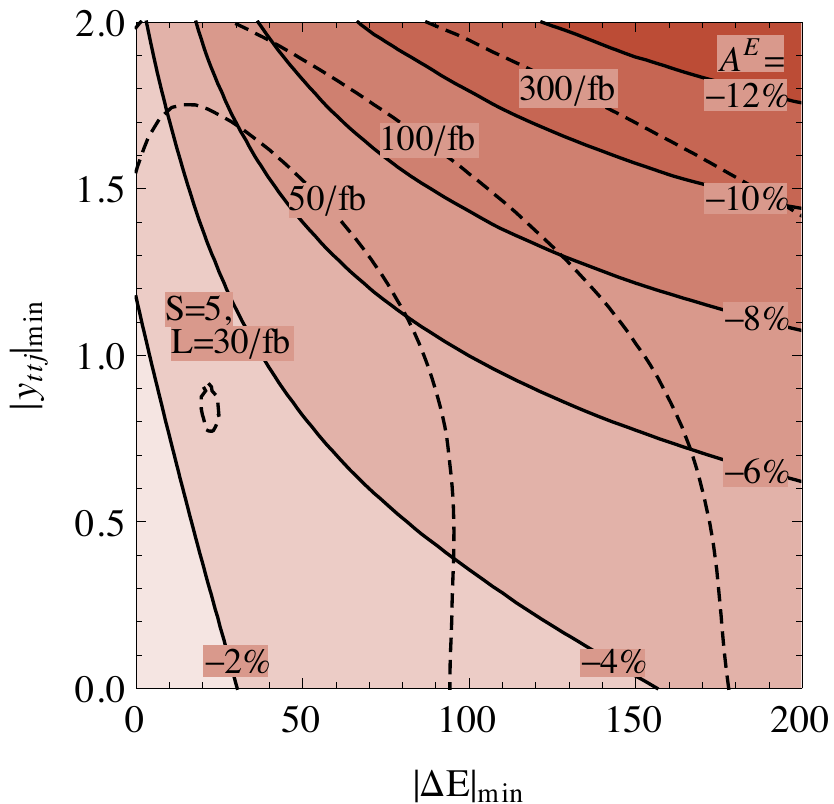}
\end{center}
\vspace*{-1cm}
\begin{center} 
\parbox{15.5cm}{\caption{\label{fig:aphi-ae-lhc14}Incline asymmetry $A^{\varphi,q}$ (left) and energy asymmetry $A^E$ (right) at LHC14, as functions of the cuts $\{|\cos\varphi|_{\text{min}},|y_{t\bar tj}|_{\text{min}}\}$ and $\{|\Delta E|_{\text{min}},|y_{t\bar tj}|_{\text{min}}\}$, respectively. An additional fixed cut on the partonic jet rapidity, $|\hat{y}_j| \le 0.5$, has been applied, as well as the detector cuts $p_T^j \ge 25\,\text{GeV}$ and $|y_j| \le 2.5$. Superimposed are contour lines of constant asymmetry (plain) and of constant statistical significance $\mathcal{S}=5$ for various luminosities (dashed).}}
\end{center}
\end{figure}

At LHC14, the maximum asymmetries are slightly reduced compared to LHC8, but still sizeable with $A^{\varphi,q} = -4\,\%$ and $A^E = -12\,\%$. The increased gluon-gluon background can be overcome by stronger cuts, once a larger data sample becomes available. 
 In Figure~\ref{fig:aphi-ae-lhc14}, we display the incline asymmetry (left panel) and the energy asymmetry (right panel) with a jet rapidity cut of $|\hat{y}_j| \le 0.5$ and variable cuts on $|y_{t\bar t j}|$, $|\cos\varphi|$ and $|\Delta E|$. The superimposed dashed lines indicate contours with a statistical significance of $5\,\sigma$ for a given integrated luminosity.
With $50\,\text{fb}^{-1}$, asymmetries of $A^{\varphi,q} = -2.4\,\%$ and $A^E = -6.5\,\%$ can be 
distinguished from the null hypothesis with a significance of $5\,\sigma$.
With a luminosity of $100\,\text{fb}^{-1}$, as projected for a three years' runtime at $14$ TeV, larger asymmetries are accessible, i.e., $A^{\varphi,q} = -3.7\,\%$ and $A^E = -8.8\,\%$. Thus, at this stage the region of a maximal incline asymmetry can be probed with appreciable precision. Digging into the kinematic region of large asymmetries is important so that the signal will not be hidden due to systematic uncertainties. With a luminosity of $300\,\text{fb}^{-1}$, an energy asymmetry of $A^E = -11\,\%$ can be observed with $5\,\sigma$ significance. These estimates show the high discovery potential for both observables at LHC14, even in the presence of sizeable systematic uncertainties.

\section{High-luminosity upgrade}
The measurement of a charge asymmetry at the LHC benefits from high luminosities in multiple respects. The gain in statistical significance speaks for itself: For a projected luminosity of $3000\,\text{fb}^{-1}$ at $14$ TeV, asymmetries of $A^{\varphi,q} = -4\,\%$ and $A^E = -11\,\%$ can be distinguished from the null hypothesis with $16$ standard deviations. Statistics-driven systematical uncertainties are expected to shrink for high luminosities, whereas enhanced pile-up effects will add to the systematical error. Event
reconstruction and the identification of the hard jet among additional
QCD radiation will be experimentally challenging. A veto on two or more
hard jets might be required in order to preserve momentum conservation
among the top pair and the hard jet.

QCD corrections may have a significant impact on the numerical
prediction of the charge asymmetry. At NLO, the cross section of $t\bar t + j$ production with subsequent decay via the dilepton channel receives a sizeable reduction \cite{Melnikov:2011qx}, calling for luminosities higher than suggested by estimates based on tree-level calculations. The effect of parton-level NLO corrections on the incline and energy asymmetries is currently under investigation \cite{Berge:2013}. Further dedicated studies of effects such as electroweak $t\bar t + j$ production, top-quark decays and parton showering are required to yield a precise prediction of the observables. High-luminosity analyses will facilitate the access to the properties of the charge asymmetry beyond inclusive observables. In this attempt, particular attention should be drawn to kinematical distributions, which reveal additional valuable information.

In view of the large forward-backward asymmetry at the Tevatron, the investigation of new-physics effects on the charge asymmetry at the LHC is of high interest. Depending on the size of the contributions to $A^{\varphi,q}$ and $A^E$, high luminosities may be required in order to disentangle new-physics effects from the standard-model prediction. The rapidity asymmetry in $t\bar t + j$ production is highly sensitive to massive color-octet bosons, which provide a solution to the Tevatron anomaly \cite{Ferrario:2009ee,Berge:2012rc}. Effects of comparable size are expected for the incline and energy asymmetries, for which a first study is in progress \cite{Stefan2:2013}. In this respect, further analyses are welcome in order to assess the sensitivity of $A^{\varphi,q}$ and $A^E$ to new-physics effects.

\section{High-energy upgrade}
\begin{figure}[!t]
\begin{center}
\includegraphics[height=8.5cm]{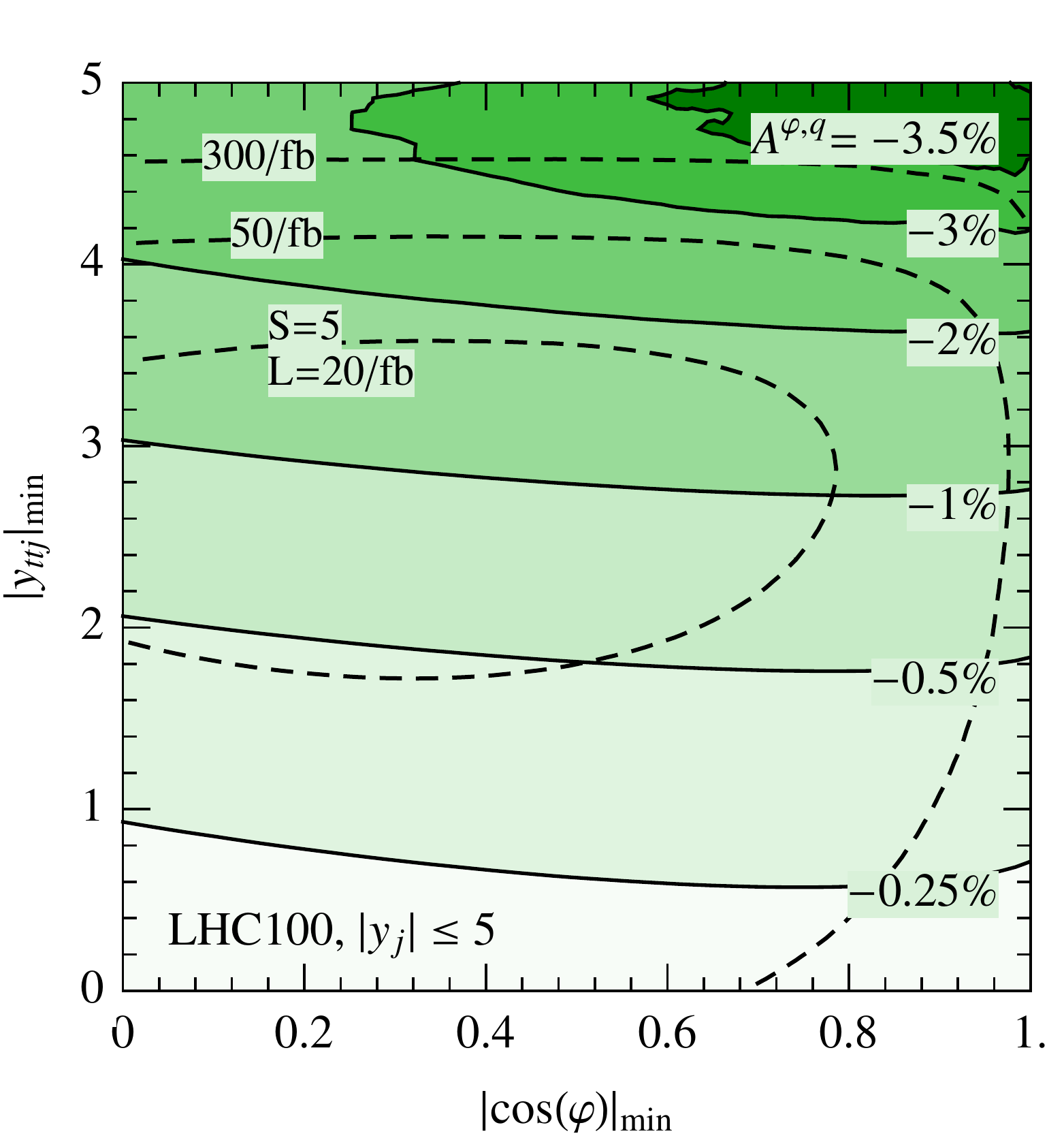}
\hspace*{0.4cm}
\includegraphics[height=8.5cm]{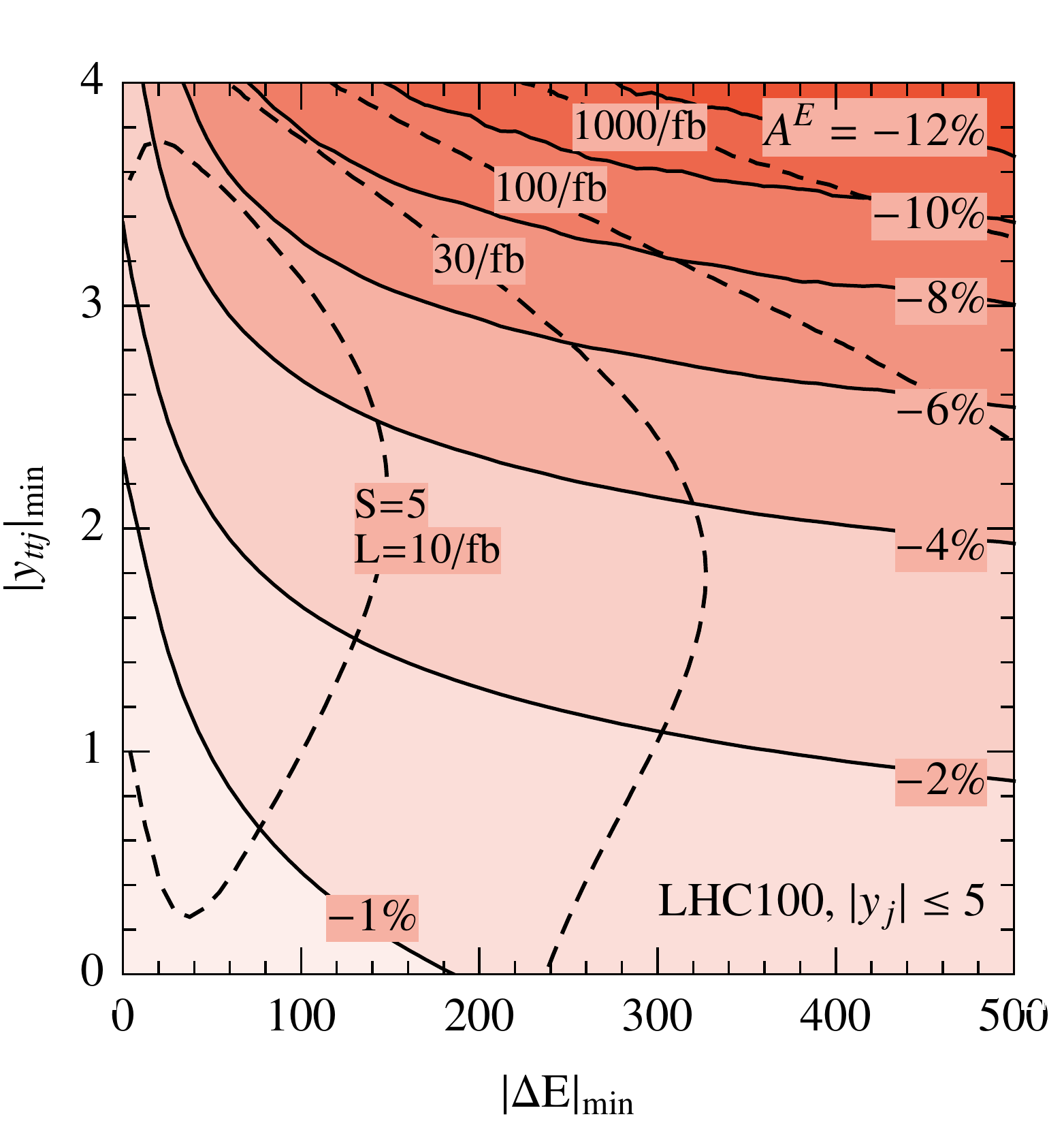}
\end{center}
\vspace*{-1cm}
\begin{center} 
  \parbox{15.5cm}{\caption{\label{fig:aphi-ae-lhc100}Incline asymmetry $A^{\varphi,q}$ (left) and energy asymmetry $A^E$ (right) at LHC100, as functions of the cuts $\{|\cos\varphi|_{\text{min}},|y_{t\bar tj}|_{\text{min}}\}$ and $\{|\Delta E|_{\text{min}},|y_{t\bar t j}|_{\text{min}}\}$, respectively. An additional fixed cut on the partonic jet rapidity, $|\hat{y}_j| \le 0.5$, has been applied, as well as the detector cuts $p_T^j \ge 25\,\text{GeV}$ and $|y_j| \le 5$. Superimposed are contour lines of constant asymmetry (plain) and of constant statistical significance $\mathcal{S}=5$ for various luminosities (dashed).}}
\end{center}
\end{figure}

At a proton-proton collider with a CM energy of $100$ TeV (LHC100), the charge asymmetry in $t\bar t + j$ production can be probed in regions of the kinematical phase space that are distinct from the region accessible at
LHC14. In Figure~\ref{fig:aphi-ae-lhc100}, we display the incline asymmetry (left panel) and the energy asymmetry (right panel) for LHC100.
As can be observed by comparing with Figure~\ref{fig:aphi-ae-lhc14}, the maximal asymmetries are almost the same as at LHC14. The gluon-gluon background at LHC100 is significantly enhanced.\footnote{The partonic contributions to the total cross section amount to $82\%$ ($gg$), $17\%$ ($qg$) and $1\%$ ($q\bar q$) at leading order for $p_{T}^j\ge 25$~GeV and $|y_j|\le 5 $.} Therefore, a stronger cut on the boost $y_{t\bar t j}$ of the final state needs to be applied in order to project on the charge-asymmetric $q\bar q$ and $qg$ contributions. The strong boost of $t\bar t + j$ events originating from initial states with quarks implies that the hard jet is preferentially emitted in the direction of the beam axis. Since most events feature jets with rapidities $|y_j| \ge 2.5$, it is indispensable to relax the detector cut up to $|y_j| \le 5$. A measurement of the charge asymmetry at LHC100 will therefore depend crucially  on the control of QCD activity close to the beam axis.

The sensitivity to $A^{\varphi,q}$ and $A^E$ at LHC100 is comparable to what has been obtained for LHC14. With $100\,\text{fb}^{-1}$, one can reach values of $A^{\varphi,q} = -3\,\%$ and $A^E = -9\,\%$ with a statistical significance of $5\,\sigma$, whereas with larger data sets, one can probe the asymmetries close to their maxima. Given these similarities, we conclude that a high-energy LHC upgrade can provide a useful cross check of the charge-asymmetry observables at LHC14.

\section{Acknowledgments}
We thank Jessie Shelton, Andreas Jung and Markus Schulze for interesting discussions and comments. This work was supported by the Initiative and Networking Fund of the Helmholtz Association, contract HA-101 (`Physics at the Terascale'), by the Research Center `Elementary Forces and Mathematical Foundations' of the Johannes Gutenberg-Universit\"at Mainz, and by the National Science Foundation, grant PHY-1212635.

\end{document}